\documentstyle[jkas]{article}

\runningauthor {Cheon et al.} \year{2010} \volume{43}
\beginpage{111}\endpage{116}
\runningtitle{Ruprecht 93}

\begin{document}

\title{No Open Cluster in the Ruprecht 93 Region}

\author{Sora Cheon$^1$, Hwankyung Sung$^1$ and M. S. Bessell$^2$}
\address{$^1$ Department of Astronomy and Space Science, Sejong University,
	98 Kunja-dong, Kwangjin-gu, Seoul 143-747, Korea \\
	{\it e-mail}: sungh@sejong.ac.kr}

\address{$^2$ Research School of Astronomy and Astrophysics, Australian
	National University, MSO, Cotter Road, Weston, ACT 2611, Australia
	{\it e-mail}: bessell@mso.anu.edu.au}

\address{\normalsize{\it (Received June 16, 2010; Accepted June 29, 2010)}}
\offprints{H. Sung}

\abstract{$UBVI$ CCD photometry has been obtained for the Ruprecht 93 (Ru 93)
region. We were unable to confirm the existence of an intermediate-age open 
cluster in Ru 93 from the spatial distribution of blue stars. On the other hand, 
we found two young star groups in the observed field: the nearer one (Ru
93 group) comprises the field young stars in the Sgr-Car arm at $d \approx 2.1$ 
kpc, while the farther one (WR 37 group) is the young stars around WR 37 
at $d \approx 4.8$ kpc. We have derived an abnormal extinction law ($R_{V}$ =
3.5) in the Ruprecht 93 region. 
}


\maketitle

\section{Introduction}

Open clusters can provide valuable information about the formation and
evolution of individual stars and stellar systems, the spiral structure of
the Galaxy, and the chemical evolution of the Galaxy. Most young open clusters
are distributed in the Galactic plane and it is therefore very difficult to
select members of the clusters due to foreground and background 
interlopers. In addition, the small number of member stars and the irregular
shape of open clusters make it difficult to test stellar evolution theory
and to compare their distribution with the results obtained from dynamical 
evolution models (Sung et al. 1999).

There are more than 1300 known objects identified as open clusters in the 
Galaxy. However many of them have still not been observed even in the $UBV$ 
system (Mermilliod \& Paunzen 2003). In addition, many have not yet been 
examined critically to see whether they are really open clusters. Ruprecht 93 
(Ru 93) was catalogued as an intermediate-age open cluster near $\eta$ 
Carinae by Ruprecht who later (Ruprecht 1966) classified the cluster as 
a Trumpler type III 2 with two chains of stars. Steppe (1977)
observed 93 stars in the cluster in the $RGU$ photographic photometric system.
They determined a distance ($d = 1.57$ kpc) and estimated the earliest
spectral type (b7) in Ru 93. The photometric study by Steppe (1977) is the only
photometric study for Ru 93 to date. The existence of Wolf-Rayet (WR) star 
WR 37 near Ru 93 could imply a younger age for Ru 93. From a photometric 
study of the Ru 93 region we have carefully examined the nature of Ru 93
and studied the possible link between Ru 93 and WR 37. 

In \S 2, we present the photometric data for 24,246 stars in the region 
of Ru 93. In addition, we have identified
the 2MASS counterpart of these stars. In \S 3 we derive the reddening law in
the region using the intrinsic color relations of early type stars in $UBVI$
and near-IR 2MASS $JHK_s$. From the distribution of reddening and the distance 
modulus of young stars, we have identified two young star groups in the 
observed region. The age of these young star groups has also been estimated.
We make some notes on WR 37, an O9 star HD 305916 in \S 4.
\S 5 is the conclusion.

\section{Observations and Data Reduction}

\begin{figure}[t]
\begin{center}
\epsfxsize=8.0cm \epsfbox{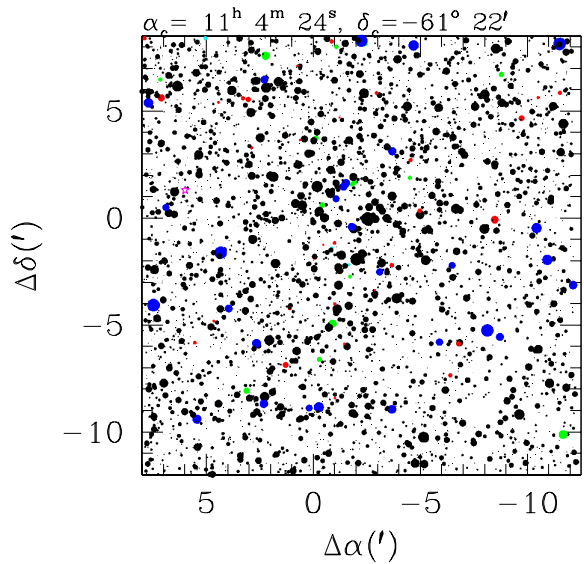}
\end{center}
\caption{Finding chart, centered at $\alpha$ = 11$^h$ 4$^m$ 24.$^s$0,
$\delta$ = -61$^\circ$ 22.$'$0 (J2000.0). The size of the symbol is 
proportional to the magnitude of the star. Blue ($E(B-V) =$ 0.2 -- 0.6 mag \&
$V_0 -M_V \leq$ 12 mag), green ($E(B-V) =$ 0.25 -- 0.8 mag \& $V_0 -M_V =$ 12 
-- 14 mag), and red-colored ($E(B-V) \geq$ 1.0 mag \& $V_0 -M_V \leq$ 14 mag)
symbols represent, respectively, the young blue stars according to their 
reddening (see \S III (b) for details). \label{map}}
\end{figure}

{\scriptsize
\begin{deluxetable}{@{}rccccccc@{}c@{}c@{}c@{}cc@{}cl@{}}
\setlength{\tabcolsep}{-1mm}
\tablecaption{Photometric Data\tablenotemark{a}}
\tablehead{
\colhead{ID} & \colhead{$\alpha (J2000.0)$} & \colhead{$\delta (J2000.0)$} & 
\colhead{$V$} & \colhead{$V-I$} & \colhead{$B-V$} & \colhead{$U-B$} &
\colhead{$\epsilon_{V}$} & \colhead{$\epsilon_{V-I}$} & 
\colhead{$\epsilon_{B-V}$} & \colhead{$\epsilon_{U-B}$} &
\colhead{n$_{obs}$} & \colhead{2MASS} & \colhead{Sp type} & \colhead{Remark}
}
\startdata
23913&11:05:31.19&-61:17:27.4& 9.253&  --  & 1.062& 0.819&0.005&  -- &0.006&0.008&1 0 1 1&11053117-6117271 &       &          \\         
10681&11:04:02.77&-61:22:02.1& 9.446&  --  & 1.667& 1.562&0.007&  -- &0.009&0.008&1 0 1 1&11040280-6122020 &       &          \\
23255&11:05:26.40&-61:26:04.2& 9.669& 0.247& 0.123&-0.684&0.005&0.013&0.009&0.008&1 1 1 1&11052637-6126041&B0.5III & HD 96355 \\
  803&11:02:48.36&-61:13:51.6& 9.812& 0.128& 0.098&-0.507&0.014&0.027&0.016&0.014&1 1 1 1&11024831-6113512 &       &          \\
 4472&11:03:15.88&-61:27:15.1& 9.853& 0.070& 0.044&-0.544&0.007&0.016&0.009&0.006&1 1 1 1&11031589-6127151&O9      & HD305916 \\
19203&11:05:00.10&-61:23:37.0& 9.882& 0.096& 0.002&-0.602&0.005&0.011&0.007&0.010&1 1 1 1&11050007-6123369 &B2Ve   & HD306016 \\
11023&11:04:05.32&-61:13:42.5& 9.977& 0.100& 0.017&-0.541&0.005&0.012&0.006&0.004&1 1 1 1&11040531-6113427&B3III   & HD306004 \\
21415&11:05:13.93&-61:20:41.3&15.126& 2.242& 1.161& 0.548&0.003&0.004&0.004&0.015&2 2 2 1&11051390-6120411&WN4     & WR 37       \\
14185&11:04:26.22&-61:25:45.8&16.303& 0.421& 0.214&-0.600&0.010&0.021&0.020&0.021&2 2 2 1&                &        & WD candidate \\
\enddata
\tablenotetext{a}{This is a sample of the full table, which is available from HS.}
\label{tab1}
\end{deluxetable}
}

$UBVI$ CCD photometry for Ru 93 was performed on 1997 June 23 using 
the 1-m telescope at Siding Spring Observatory. During the observing run 
we observed SAAO E region standards (E5 and E7). The standard transformation 
relations are discussed in Sung \& Bessell (2000), however some coefficients
have been modified to avoid the small, non-linear corrections found in 
the $I$, $B$ and $U$ transformations. The details of these modifications
will be discussed in a forthcoming
paper. The atmospheric extinction coefficients were 0.069 ($\pm$ 0.009),
0.141 ($\pm$ 0.014), 0.277 ($\pm$ 0.009), and 0.520 ($\pm$ 0.016) in $I$,
$V$, $B$, and $U$, respectively. The secondary extinction coefficients
in $B$ and $U$ from Sung \& Bessell (2000) have been adopted and used.

The two sets of exposure times for Ru 93 - 30s and 600s in $U$, 10s and 300s in
$B$, 5s and 120s in $V$, and 5s and 60s in $I$ - were used to secure both 
bright and faint stars in the observed field. The point spread function fitting
photometric package DAOPHOT implanted in IRAF was used. A total of 24,246 stars
were observed in the region. The only saturated star was a K5 star HD 96193
($V$ = 7.39). Figure \ref{map} is the finding chart of the Ru 93 region.
In order to determine the reddening law in Ru 93, we have identified the 2MASS
counterpart of optical sources. Table \ref{tab1} shows photometric data
for several bright stars, WR 37, and a white dwarf candidate. We used different
colors for different groups in Figure \ref{map} (see \S 3 (b) for 
the classification of young star groups). We can easily see that there is 
no clustering of blue stars in the figure. In addition, there is no enhancement
of bright blue stars in the presumed core of Ru 93 near ($\Delta \alpha$ = 0.0,
$\Delta \delta$ = 0.0). This fact implies that either 
there is no open cluster or if there is an open cluster its size  
is larger than the field of view of our CCD.

\section{Photometric Diagrams and Young Star Groups in the Ruprecht 93 Region}

\subsection{Reddening Law}

\begin{figure*}
\begin{center}
\epsfxsize=16.0cm \epsfbox{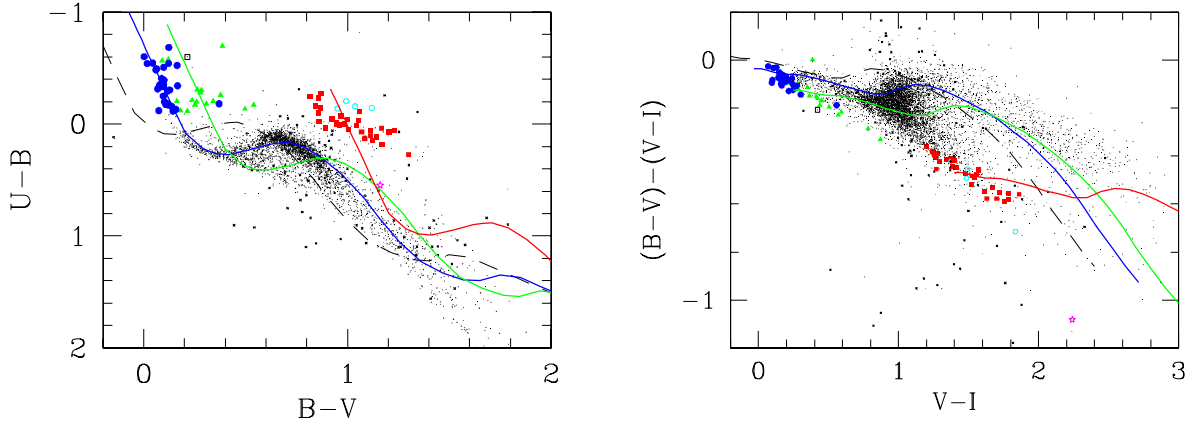}
\end{center}
\caption{Color-Color Diagrams. The thin dashed line represents the ZAMS 
relation, while the three solid lines represent the ZAMS relations 
reddened by $E(B-V)$ = 0.25, 0.44, and 1.25 from left to right, respectively. (Blue) Filled 
circles, (green) triangles, and (red) squares are the stars belonging to these 
three groups. The star mark (magenta) denotes WR 37, and a dot within a square 
represents the white dwarf candidate ID 14185. The open circles at $B-V 
\approx$ 1.0 show slight differences in photometric characteristics
(see \S III (b) for details). Small dots and crosses represent, respectively, 
good data ($\epsilon < 0.1$) and bad data ($\epsilon \geq 0.1$). \label{ccd} }
\end{figure*}

We present the color-color diagrams from our photometry in Figure \ref{ccd}. 
It is very easy to discern that there are at least two groups of reddened blue 
stars. The median $E(B-V)$ of the least reddened blue stars (filled circles)
is about 0.25 mag, and that of the highly reddened stars (squares) is 
about 1.25 mag. Several blue stars can be found between these two (triangles). 
WR 37 (star mark) is fairly red in $(U-B)$. Since there is no 
intrinsic color relation for WR stars, especially in the broad-band standard 
photometric systems, due to the contamination by broad emission lines, 
it is thus impossible to determine
the reddening of WR 37. But the colors of WR 37 strongly imply 
that WR 37 belongs to the more highly reddened group. We also present
the [$(B-V)-(V-I)$, $(V-I)$] diagram in the right panel of Figure \ref{ccd}.

\begin{figure}
\begin{center}
\epsfxsize=8.5cm \epsfbox{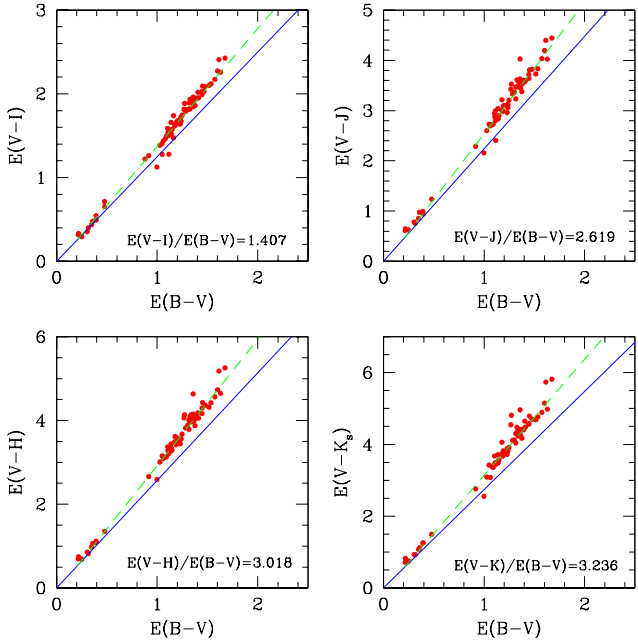}
\end{center}
\caption{Color excess ratios. The color excess of each color is calculated
using the relation between intrinsic colors. See text for details.
\label{red_law} }
\end{figure}

Knowledge of the reddening law, especially the total-to-selective 
extinction ratio $R_V \equiv A_V / E(B-V)$, is very important in estimating
the distance to an object. In general, $R_V$ can be determined from the color
excess ratio (Guetter \& Vrba 1989). The reddening $E(B-V)$ of individual 
early type stars is calculated from the color-color diagram shown in Figure 
\ref{ccd}. The color excess of each color is calculated using the relation
between intrinsic colors. The intrinsic color relation between $B-V$ and
$V-I$ is taken from Sung \& Bessell (1999). Those for 2MASS $JHK_s$ were calculated
by one of the authors (MSB). 

Figure \ref{red_law} shows the relation between $E(B-V)$ and color excesses in
other colors. The thick solid line represents the color excess relation for the
normal case, i.e. $R_V = 3.1$. Evidently early type stars in the Ru 93 region
show slightly abnormal values. When we adopt the foreground reddening of
$E(B-V)_{fg} = 0.22$, the median value for $E(V-I)_{\rm Ru ~93} / E(B-V
)_{\rm Ru 93}$, $E(V-J)_{\rm Ru ~93} / E(B-V)_{\rm Ru ~93}$, $E(V-H)_{\rm 
Ru ~93} / E(B-V)_{\rm Ru ~93}$, and $E(V-K_s)_{\rm Ru ~93} / E(B-V)_{\rm 
Ru ~93}$ is 1.407, 2.619, 3.018, and 3,238, respectively, where
$E(V-C)_{\rm Ru 93}$ is defined by

$$E(V-C)_{\rm Ru 93} \equiv E(V-C) - E(V-C)_{fg} $$ 
$$= E(V-C) - [{{E(V-C)}\over{E(B-V)}}]_{fg}\cdot E(B-V)_{fg}, $$

\noindent
($C$ is $I$, $J$, $H$, or $K_s$).
We adopt the color excess ratios for the foreground
field $[E(V-C)/E(B-V)]_{fg}$ of 1.25, 2.24, 2.57, and 2.74 for $I$, 
$J$, $H$, and $K_s$, respectively. These values are very similar to those
of a $R_V = 3.1$ star HD 167771 (Fitzpatrick 1999). If we adopt the relation 
between the color excess ratio and $R_V$ of Guetter \& Vrba (1989), the color 
excess ratios give $R_{V} = 3.51 \pm 0.05$ and thus we adopt $R_{V,{\rm 
Ru ~93}} = 3.5$ for the Ru 93 region.\footnote{If we adopt the relation between 
the color excess ratio and $R_V$ of Fitzpatrick (1999) (see their equations
A3, A4, and A5), the resulting $R_V$ is $3.63 \pm 0.01$.}

\subsection{Two Young Star Groups}

\begin{figure*}
\begin{center}
\epsfxsize=16.0cm \epsfbox{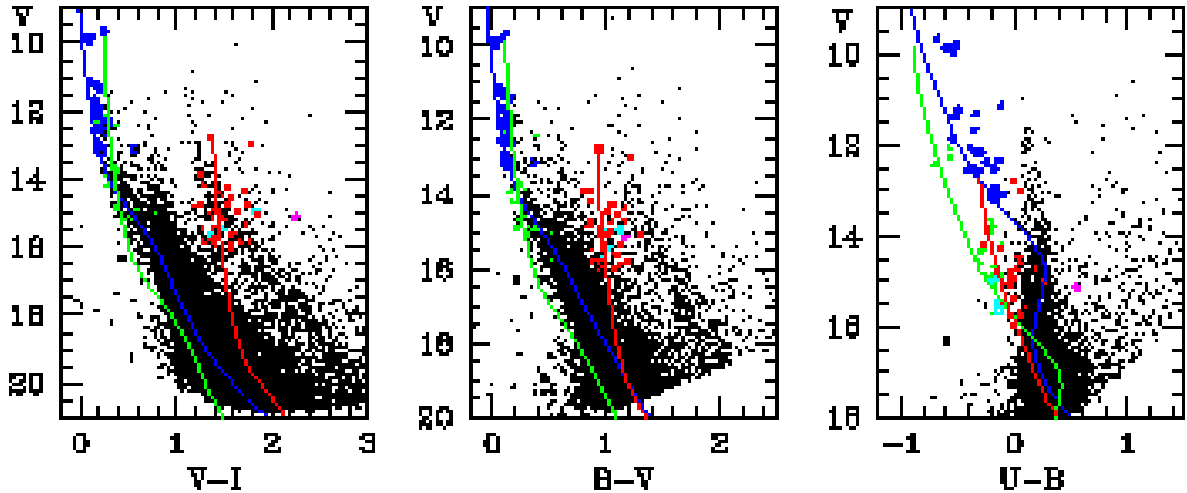}
\end{center}
\caption{Color-Magnitude Diagrams. Three solid lines represent the reddened 
ZAMS relations reddened by $E(B-V)$ = 0.25 and $V_0 - M_V = 11.7$, $E(B-V)$ = 
0.44 and $V_0 - M_V = 13.4$, and $E(B-V)$ = 1.25 and $V_0 - M_V = 13.4$, 
respectively. All symbols are the same as Figure \ref{ccd}. \label{cmd} }
\end{figure*}

\begin{figure}
\begin{center}
\epsfxsize=8.5cm \epsfbox{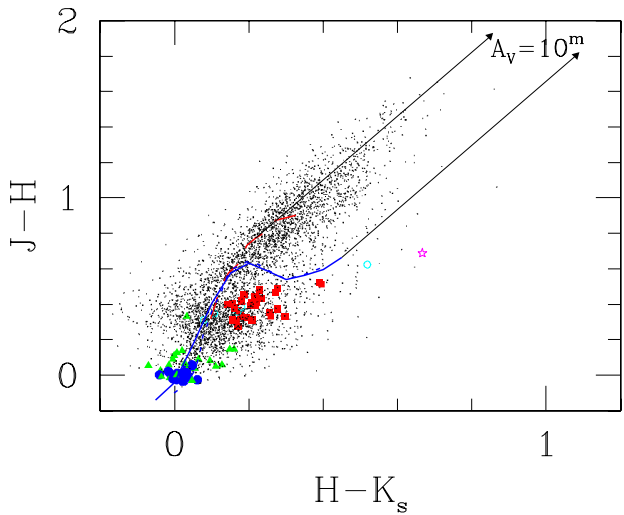}
\end{center}
\caption{Near-IR $JHK_s$ color-color diagram for the stars with 
$\epsilon_{J,H,K_s} \leq 0.1$. Two dashed lines are the unreddened relation 
for MS stars (short dashed line) and giants (long dashed line) from Bessell 
\& Brett (1988), and the solid line is the unreddened MS relation in 2MASS
$JHK_s$ from Sung et al. (2008). The thin lines show the direction of the
reddening vector of $A_V = 10$, and the stars between these two lines are
reddened normal stars. The other symbols are the same as Figure
\ref{ccd}. The stars ID 20126 (open circle) and WR 37 show an IR excess. 
\label{nirccd} }
\end{figure}

The distance to an open cluster can be determined using the relation between
reddening corrected colors and the absolute magnitudes of zero age main sequence
(ZAMS) stars. As mentioned in \S 2, there is no clustering of blue stars
in the observed region. This fact implies that it is practically impossible
to determine the distance to these stars. Nevertheless, the color-magnitude 
diagram (CMD) of these blue stars in Figure \ref{cmd} is very similar to 
that of an open cluster. That is because young blue stars can only be in spiral
arms as members of OB associations or young open clusters.
We calculated the distances to the three blue star groups in Figure \ref{ccd}.
The distance modulus of the less reddened group is about 11.6 mag ($d = 2.1$ 
kpc). Interestingly, the distance modulus of the other two groups is very 
similar.\footnote{Another
interpretation for the intermediate reddened group is that they could be 
pre-main sequence (Herbig Ae/Be) stars associated with the less reddened group. 
But their numbers are too many and the amount of UV excess is too large 
when we compare their $(U-B)$ in Figure \ref{ccd} with the Herbig Ae/Be 
star W90 in NGC 2264 (Sung et al. 1997).} We adopt the distance modulus to 
the redder groups as $V_0 - M_V = 13.4$ ($d = 4.8$ kpc). We present the CMDs 
in Figure \ref{cmd}. The reddened ZAMS relations with an appropriate distance 
modulus are shown in the figure. We call the less reddened group the Ru 93 
group, and the more highly reddened group the WR 37 group.

From Figure \ref{ccd} it is seen that the four apparently highly reddened 
objects having slightly different photometric characteristics. Their distance 
modulus is further away than the reddened group ($V_0 - M_V \approx 14.5$, 
i.e. $d \approx 7.9$ kpc). But it is also possible to interpret their 
$(U-B)$ color as being an UV excess as seen in many classical T Tauri stars. 
We checked their colors carefully and found that one of them (ID 20126 = 
2MASS J11050589-6113350) shows an evident IR excess (see Figure \ref{nirccd}).
In addition the star shows excess emission in the $(B-V)-(V-I)$ versus $(V-I)$
diagram in Figure \ref{ccd}. The star ID 20126 may be a pre-main sequence star
in the less reddened group. However, the other three stars do not show any 
signature of excess emission in the near-IR $JHK_s$ or ultraviolet.

The distances to these young star groups are related to the spiral arm structure
of the Galaxy. When we compare their location ($l = 290.^\circ47,~ 
b=-1.^\circ11$) in the galactic plane with the distribution of HII regions 
and giant molecular clouds (see for example Fig. 5 of Hou et al. 2009), 
their connection with the spiral arms is seen. At $l \approx 290^\circ$, 
the line of sight passes the inner arm twice
- once at $d \approx 2.1$ kpc and the other at $ \approx 7$ kpc. Two HII
regions and a giant molecular cloud can be found between them. The less 
reddened group is coincident with the near side of the inner spiral arm. 
The highly reddened group is not coincident with the inner spiral arm but
is between the two crossings. It is worth noting however, that the width 
of a spiral
arm is about 500 pc and in addition, the distribution of spiral arm tracers is
not well matched with the spiral arm structure of the Galaxy (see Figure 5
of Hou et al. 2009).

\subsection{Age of Young Star Groups}

\begin{figure}[t]
\begin{center}
\epsfxsize=8.5cm \epsfbox{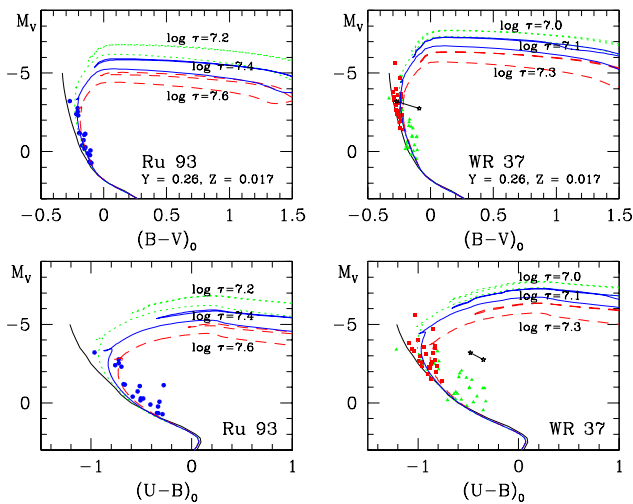}
\end{center}
\caption{H-R diagrams of the Ru 93 and WR 37 groups. The superimposed lines are
the isochrones from Bertelli et al. (2009) for solar abundances. The star
marks represent the position of WR 37 for $E(B-V)$ = 1.25 (the median
$E(B-V)$ of WR 37 group) and $E(B-V)$ = 1.43 from van der Hurst (2001). 
\label{hrd} }
\end{figure}

As there is no evident clustering of stars in the observed region we cannot
derive an age from cluster fitting. However, although there is no known 
OB association 
(Humphreys 1978) in the observed region, the blue stars may comprise part 
of unknown OB associations. The age of young blue stars in the region could
represent the age of the youngest population in this region of the spiral arm.

Figure \ref{hrd} shows the observational Hertzsprung-Russell (H-R) diagrams
of the Ru 93 and WR 37 groups. Superimposed are the most recent isochrones 
from the Padua group (Bertelli et al. 2009) for solar abundances (Y=0.26, Z=0.017). 
As the solar abundance models of the Padua group are brighter by about 0.3 mag
(Sung et al. 1999), the isochrones were shifted vertically by 0.3 mag.
The age of the Ru 93 group is between $\log \tau =$ 7.2 and 7.6, which is 
slightly younger than the Pleiades. The age of the WR 37 group is, on the other 
hand, much younger than the Ru 93 group ($\log \tau \leq 7.2$). Several stars 
in the WR 37 group are smaller than the age of the youngest isochrone of 
Bertelli et al. (2009). If we fit these blue stars with the isochrones of 
Bertelli et al. (1994) the age of the youngest stars is given by $\log 
\tau = 6.6$. This implies that the star formation in the region around 
WR 37 occurred very recently (about 4 Myr ago).

\section{Individual Stars}

\subsection{WR 37}

One of main purposes of this study was to estimate the mass of WR 37. As we
could not find any young open cluster associated with WR 37, we were unable to
constrain the mass and the evolutionary status of WR 37 from cluster age fitting.

As there is no known intrinsic color relation for WR stars in the broad-band 
photometric system, we cannot estimate the reddening and the intrinsic colors 
of WR 37. After applying the mean value of reddening, the intrinsic colors
of WR 37 are $(V-I)_0$ = 0.46, $(B-V)_0$ = -0.09, $(U-B)_0$ = 0.26, and $M_V$ 
= -2.75. Van der Hurst (2001) estimated the reddening to WR 37 as $E_{b-v}$ 
= 1.43 from narrow-band photometry, then WR 37 becomes slightly bluer and 
brighter ($(V-I)_0$ = 0.24, $(B-V)_0$ = -0.27, $(U-B)_0$ = -0.48, and $M_V$ 
= -3.19). He also estimated the distance to WR 37, which is the same value that
we obtained for the WR 37 group ($V_0 - M_v$ = 13.4).

\subsection{HD 305916 (=ID 4472)}

HD 305916 (= CPD -60 2483) is classified as O9V (Cruz-Gonzalez et al. 1974).
The estimated reddening $E(B-V)$ = 0.25, and the distance modulus using the ZAMS
relation is 9.7.

If the spectral type of HD 305916 is correct, then the distance modulus of 
the star becomes 13.5 ($M_V$ (O9V) = -4.4: Sung 1995). In addition, 
the reddening corrected color of the star should be bluer. If HD 305916 
belongs to the WR 37 group, the star is much brighter in Figure \ref{hrd}. 
The reddening corrected colors are very similar to those of B type stars. 
The spectral type of this star needs to be re-examined.

\section{Conclusions}

$UBVI$ CCD photometry has been obtained for the region around WR 37, which was
previously considered to contain the intermediate-age open cluster Ru 93. 
Using optical photometry as well as near-IR 2MASS data, we found an abnormal 
extinction law in the region ($R_{V,~{\rm Ru~93}}$ = 3.5) and identified two or 
more groups of blue stars along the line of sight. But we could not find any 
concentration of blue stars in the observed region. We conclude that Ru 93 
is not a real open cluster. 

We have determined the distance to the blue stars, and found that the less
reddened group of blue stars (Ru 93 group) is at about $2.1$ kpc which 
coincides with the closest part of the inner arm. The distance to the more 
highly reddened group (WR 37 group) is about $4.8$ kpc, which is intermediate 
between two crossings of the inner spiral arm along the line of sight, 
and therefore is not well matched with the spiral arm structure of the Galaxy. 
The age of the young stars in the Ru 93 group ($\log \tau =$ 7.2 -- 7.6)
is slightly larger than the young stars in the WR 37 group ($\log \tau 
\leq 7.2$).

\acknowledgments

H. S. acknowledges the support of the National Research Foundation of Korea
(NRF) to the Astrophysical Research Center for the Structure and Evolution 
of the Cosmos (ARCSEC$''$) at Sejong University (NRF No. 2009-0062865).

\end{document}